\documentclass[aps,prl,twocolumn,superscriptaddress]{revtex4}
\usepackage{graphicx}

\begin{document}

\title{Quantum Phase Transition in the SU(4) Spin-Orbital Model
on the Triangular Lattice}

\author{Karlo Penc}
\affiliation{Max-Planck-Insitut f\"ur Physik komplexer Systeme, 
N\"othnitzer Str. 38, 01187 Dresden, Germany} 
\affiliation{Research Institute for Solid State Physics and Optics, 
H-1525 Budapest, P.O.B. 49, Hungary}

\author{Matthieu Mambrini}
\affiliation{Laboratoire de Physique Quantique, Universit\'e Paul Sabatier, 
31062 Toulouse Cedex 04, France}

\author{Patrik Fazekas}
\affiliation{Research Institute for Solid State Physics and Optics, 
H-1525 Budapest, P.O.B. 49, Hungary}

\author{Fr\'ed\'eric Mila}
\affiliation{
Institut de Physique Th\'eorique, Universit\'e de Lausanne, 
CH-1015 Lausanne, Switzerland}
\date{\today}

\begin{abstract}
Motivated by the absence of cooperative Jahn-Teller effect in
LiNiO$_2$ and BaVS$_3$, two layered oxides with triangular 
planes, we study the SU(4) symmetric spin-orbital model on the 
triangular lattice. Upon reducing the next-nearest neighbour 
coupling, we show that the system undergoes a quantum phase
transition to a liquid phase. A variational approach to this liquid
phase shows that simple types of long-range correlations are
suppressed, suggesting that it is stable against lattice distortions.
\end{abstract}


\maketitle
The possibility to enhance quantum fluctuations and 
destroy magnetic long-range order by orbital 
fluctuations was first proposed a few years ago in the pioneering
work of Feiner {\it et al} \cite{feiner}. A careful analysis of the 
spin-orbital model of cubic perovskites with two-fold orbital
degeneracy revealed the presence of spin liquid phases in which 
orbital degrees of freedom form valence-bond patterns.
The search for experimental realizations of this intriguing
physics has been going on since then, but the best candidate
remains the old system 
LiNiO$_2$ \cite{linio}, a layered S=1/2 system which does not 
undergo any phase transition down to the lowest temperatures. 
In this system, Ni$^{3+}$ ions are arranged in triangular layers, and
since the SU(2) Heisenberg model on a triangular lattice
is now believed to sustain long-range order \cite{QDJS}, the origin 
of this behaviour has logically been attributed to the two-fold
orbital degeneracy of Ni$^{3+}$ in the low spin state. 
However, the absence of any structural distortions, which one would expect 
e.g. from a valence bond-ordered spin liquid of Ref.~\cite{feiner}, 
is still a 
mystery. Similar physics seems to be present in
BaVS$_3$ \cite{bavs}, a $3d^1$ system with stacked triangular planes: Between
30K and 70K it develops a spin gapped phase, again without significant lattice
distortions.  We believe that 
essential aspects of this intermediate phase are also describable 
in terms of a triangular spin-orbital model with twofold orbital degeneracy.

On the theory side, it has become clear since the work of Feiner
{\it et al} \cite{feiner} that states in which orbitals form 
valence-bond dimer patterns do not exhaust the possibilities of 
spin liquid states. In particular, an increasing attention 
has been devoted to the SU(4) symmetric Kugel-Khomskii model \cite{kk}, 
where the orbital and spin degrees of freedom
play symmetric roles. In terms of spin-1/2 operators 
$\vec \sigma_i$ for the spins and pseudo-spin 1/2 operators $\vec \tau_i$ 
for the orbitals, this model is defined by the Hamiltonian:
\begin{equation}
  {\cal H } = \sum_{(i,j)} J_{ij} (2 \vec \sigma_i.\vec \sigma_j + \frac{1}{2})
(2 \vec \tau_i.\vec \tau_j + \frac{1}{2})
\label{KK}
\end{equation}
This Hamiltonian can be rewritten as
$  {\cal H } = \sum_{(i,j)} J_{ij} P_{i,j} $,
where $P_{i,j}$ permutes the states of the four-dimensional fundamental
representations of SU(4) at sites $i$ and $j$.
In 1D, this model has a gapless spectrum with a fourfold periodicity
of correlation functions \cite{1Dsu4}.
The investigation of this model on other simple lattices has revealed 
that, whenever possible, there is a tendency to form four-site 
plaquettes to accommodate SU(4) singlets \cite{lima}, in contrast
to the orbital valence bond solids. 
These plaquettes have been unambiguously 
shown to form a well defined pattern in the case of ladders \cite{vdb},
and evidence that this might also be the case for the square
lattice has been put forward \cite{RVBsu4,fermiMF}. Alternatively, 
long range
order has been proposed in Ref.~\cite{shen}. 
It is interesting to note that this SU(4) model is also a good 
starting point to describe CeB$_6$
(e.g. Ref.~\cite{thalmeier}), while the other version 
with conjugate representations on different sublattices is more
appropriate in the 
context of strong coupling to the lattice \cite{lattice}, or to perform
a $1/N$ expansion \cite{sachdev}.

In this Letter, we present strong evidence, all based on arguments
that respect the underlying SU(4) symmetry, that the SU(4) model on 
a triangular lattice with nearest-neighbour (NN) exchange $J>0$ only is 
also a spin liquid, but unlike previous
suggestions \cite{lima,fermiMF}, we put forward detailed variational
arguments that go against the presence of plaquette order in the ground
state. Extending the model by including a coupling $J'>0$ between 
next-nearest neighbours (NNN), we prove that the model undergoes a quantum
phase transition between a disordered state for $J'=0$ and an ordered state 
for large enough $J'$. 
The emerging picture of the disordered phase 
is that of a true liquid both in spin
and orbital sector, with accordingly no tendency to undergo any
distortion when coupled to the lattice. 

\begin{figure}
\includegraphics[width=7.0truecm]{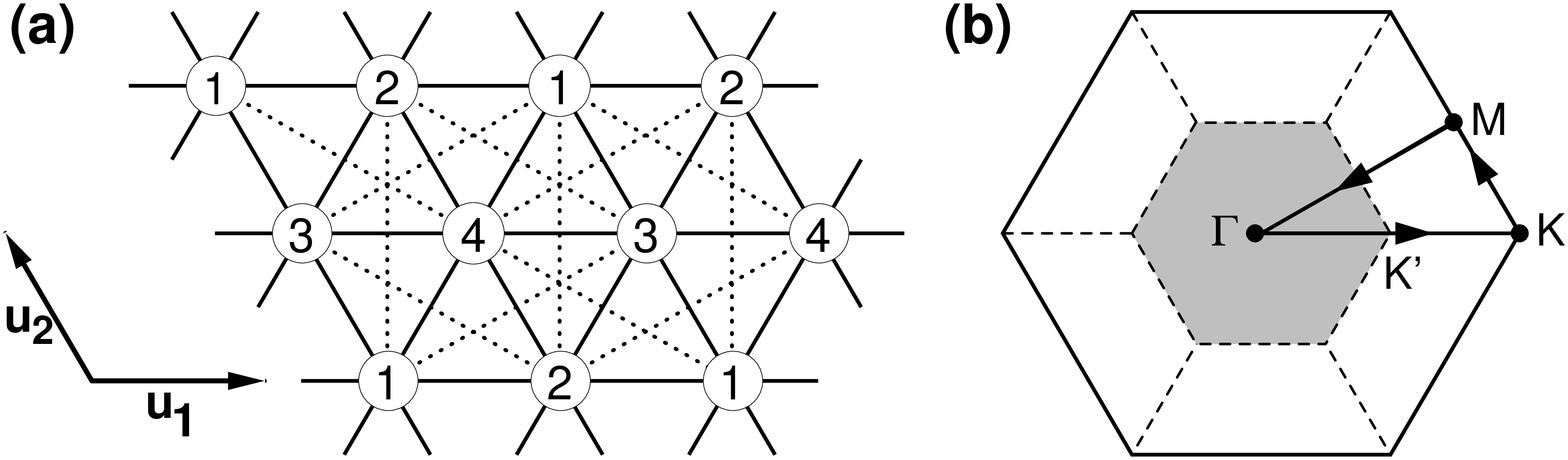}
\caption{\label{fig:BZ} (a) Four-sublattice long-range ordered
state with basis vectors ${\bf u}_1=(1,0)$ and ${\bf u}_2=(-1/2,\sqrt{3}/2)$.
The dashed line represents the next-nearest neighbor exchange. The sublattices
are labelled according to the occupation by the states of the fundamental 
representation.
(b) Brillouin-zone of the triangular lattice. The smaller, shaded hexagon
 denotes the reduced Brillouin-zone associated with the 4-sublattice order.} 
\end{figure}
  
{\it Spin-waves.}--The infinite degeneracy of the antiferromagnetic, 
four-state Potts model on the triangular lattice is lifted by $J'>0$,
leading to a unique, four-sublattice ordering pattern 
(see Fig.~\ref{fig:BZ}(a)). 
Once the classical state is ordered, quantum fluctuations can be
analyzed within a flavor-wave theory \cite{spinwave}, which is a
generalization of the standard SU(2) linearized spin-wave theory.  First, one extends the
fundamental representation on a single site into a Young diagram
with a single row and M columns (the fully symmetric
representation). The SU(4) operators can be conveniently
expressed by a generalized Holstein-Primakoff transformation,
where the fluctuations on each site are described by three different 
bosons $b_n^m(l)$ with $m\neq n$, for which the vacuum state is  
the $M$-fold direct product of the state $|m\rangle$ associated with the
particular sublattice (Fig.~\ref{fig:BZ}(a)). 
Following the standard procedure \cite{spinwave}, i.e. 
expanding in $1/M$ and keeping the quadratic bosonic terms, the
Hamiltonian can be diagonalized and, up to a constant, 
 is given by:
\begin{equation}
 \mathcal{H}_{\rm FW} = 
  \sum_{{\bf k}\in {\rm RBZ}} \sum_{m,n} 
  \omega_{mn}({\bf k}) 
 \left[ 
  \alpha^{n\dagger}_{m,{\bf k}} \alpha^{n}_{m,{\bf k}} +\frac{1}{2}
 \right] 
 \;
\end{equation}
The energies of the flavour modes are given by
\begin{equation}
  \omega_{mn}(k) = 2\sqrt{(J+J')^2 -[J \cos({\bf k}{\bf r}_{mn}) +J'
\cos({\bf k}{\bf r}'_{mn})]^2 }
\end{equation}
where ${\bf r}_{mn}$ and ${\bf r}'_{mn}$ are the NN and NNN distances between
sublattices $m$ and $n$. There are 12 branches in the reduced
Brillouin zone (RBZ, Fig.~\ref{fig:BZ}(b)), which is equivalent to 3 in the 
normal Brillouin zone. 
The operators $\alpha$ and $b$ are related through a Bogoljubov transformation.

\begin{figure}[b]
 \includegraphics[width=8.0truecm]{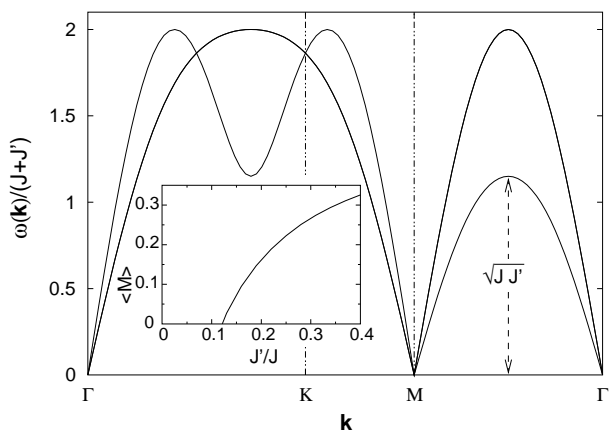}
 \caption{\label{fig:fw_disp} Flavor-wave dispersion for $J'/J=0.1$. 
Inset: Reduced moment as a function of $J'/J$.}
\end{figure}

In Fig.~\ref{fig:fw_disp} we show a plot of the flavour dispersion for some 
arbitrary value of $J'/J$.
For $J'=0$, the energy of the flavour excitations becomes effectively
one-dimensional \cite{spinwave} 
(e.g.  $\omega_{12}({\bf k})=2 J |\sin k_x|$, with zero energy
node along the $k_x=0$ line). These low energy quantum fluctuations 
destroy the 4-sublattice long-range
order. Including $J'$, the magnons along the $\Gamma-M$ line acquire 
finite dispersion $\propto \sqrt{JJ'}$, and this will reduce the 
fluctuations and  stabilize the LRO state. 
Quantitatively the fluctuations are reflected in the effective staggered 
moment $\langle M \rangle = 
M - \langle  \sum_{i \neq m} b^{m\dagger}_{i}(l)
b^{m}_{i}(l) \rangle$, which reads
\begin{eqnarray}
  \langle M  \rangle &=& M -
  \sum_{n\neq m} \left\langle \frac{J+J'}{\omega_{mn}({\bf k})}-\frac{1}{2}
  \right\rangle_{\rm RBZ}\nonumber \\
 &\approx& M + \frac{3}{2} 
  + \frac{3}{2 \pi} \ln \frac{J'}{16 J}
 + \mathcal{O}(J'/J) \;.
\end{eqnarray}
The logarithmic divergence is the manifestation of the low energy modes.
The 
$\langle M \rangle $ for $M=1$ is finite when $J'/J \gtrsim 0.12$ 
(inset in Fig.~\ref{fig:fw_disp}), and
we take this criterium for the stability of the LRO. 
Note that, within the linear flavor wave theory, the staggered 
moments and the energy are symmetric in $J$ and $J'$. So 
we expect LRO to be stable for $0.12 \lesssim J'/J \lesssim 8$ 
(for $J=0$ we have four decoupled intercalated triangular lattices).

{\it Tower of states.}-- To assess the validity of these results in the 
extreme quantum limit of the fundamental representation, we have looked
for signatures of the quantum phase transition in the exact spectrum on
a finite system. Specifically, we have numerically diagonalized 
a 12-site cluster defined by the lattice vectors 
${\bf T}_1=2 {\bf u}_1+4 {\bf u}_2$ and 
${\bf T}_2=4 {\bf u}_1+2 {\bf u}_2$. This particular cluster inherits  
the $D_6$ symmetry of the triangular lattice, and it is also compatible 
with the 4-sublattice ordering. Our aim was to identify the existence 
of the so-called Anderson tower spectrum, a set of states well separated from
the others, which become degenerate and give rise
to a linear combination with LRO in the thermodynamic limit
\cite{QDJS,andersontower}. 
For SU(2) models, their energy for an $N$-site system, measured from the 
ground state, is proportional to ${\bf S}^2/N$, where the total spin 
$S$ is the addition of sublattice spins which maximize the sublattice 
magnetizations \cite{andersontower2}.   
In the SU(4) case, the same formula is expected to hold
if ${\bf S}^2$ is replaced by the first Casimir operator $C$ of
the SU(4) algebra (see Ref.~\cite{RVBsu4} for definition and convention).

\begin{figure}[t]
 \includegraphics[width=7.0truecm]{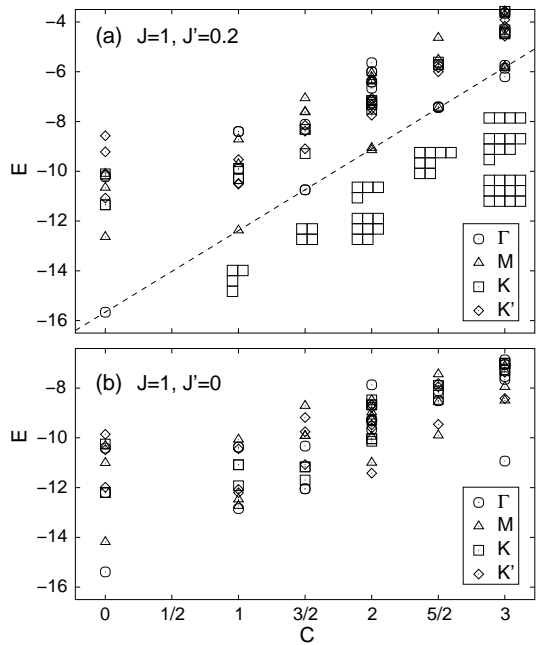}
 \caption{\label{fig:ED} Low-energy spectrum of the 12-site cluster as
a function of the first Casimir $C$ for $J'/J=0.2$ (a) and $J'=0$ (b). 
The Anderson tower of states is clearly identified
 for $J'/J=0.2$ (upper plot). 
The Young-diagrams corresponding to the states that build up the tower
are depicted.}
\end{figure}

These states can be easily identified in Fig.~\ref{fig:ED}(a),
which shows the spectrum for $J'/J=0.2$. In particular, 
the degeneracies and the symmetry properties 
are in complete agreement with group theoretical calculations \cite{further}.
This structure is also present in the spectra obtained for   
for all values of $J'/J$ larger than $0.2$ up to $J'=J$, 
including $J'=J/2$, a special
point for the 12-site cluster where the tower structure can be
explicitly proven \cite{further}.
By contrast, the spectrum for $J'=0$ is shown in Fig.~\ref{fig:ED}(b). 
Clearly, the tower structure is lost: 
The excitation energies are not aligned on a straight line, and the
symmetry and degeneracy of the corresponding excited states do not
agree with group theory. This behaviour is consistent with a
quantum phase transition between $J'/J=0$ and $0.2$. Due to the very
rapid increase of the size of the Hilbert space, we could not study
larger systems to check that the tower structure is still there for
large enough $J'/J$, and to perform a finite-size scaling of the slope.
Still, the comparison of the spin-wave results, obtained in the 
thermodynamic limit, with these exact diagonalization results gives a
very strong indication that there is a quantum phase transition
between a four-sublattice ordered state for large enough $J'/J$ and 
some kind of spin liquid for $J'=0$.

\begin{figure}[b]
 \includegraphics[width=6.0truecm]{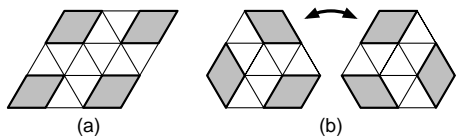}
 \caption{\label{fig:covering} Three remarkable covering patterns 
of the triangular lattice with plaquettes. The last two resonate strongly.}
\end{figure}

{\it Variational approach.}-- 
What is the nature of this 
disordered ground state? For SU(2) models, experience with frustrated systems
has shown that it is useful to look at
the structure of the low-lying singlet spectrum of a finite
system, which can contain
a single state in valence bond solids with no lattice-symmetry breaking,
four states for true spin liquids on a torus 
(topological degeneracy \cite{topdeg}), 
a finite number of states in valence bond solids with 
lattice-symmetry breaking, or a proliferation 
of low-lying singlets, as
in the spin-1/2 kagome antiferromagnet\cite{lhuillier}. In the present case,
exact diagonalization results on finite-size clusters with 12 and 16 sites
show that there are 4 (resp. 6) singlets before the first multiplet,
but no clear tendency can be identified.

\begin{table}[t]
\begin{tabular}{cccccccc}
\hline
 $N$ & $E$ diag.  &0 step  & 1 step  & 2 steps  & Exact & deg. & $k$    \\
\hline
12 & -9.75 & -14.657 & -15.314 & -15.381 & -15.384 & 1 & $\Gamma$ \\
12 & -9.75 & -12.206 & -13.781 & -14.141 & -14.188 & 3 & $M$ \\ 
\hline
16 & -13 & -19.253 & -20.935 & -21.068 & -21.079 & 1 & $\Gamma$ \\
16 & -13 & -15.767 & -17.979 & -18.634 & -18.908 & 2 & $\Gamma$ \\
16 & -13 & -16.798 & -18.252 & -18.624 & -18.754 & 3 & $M$ \\
\hline
\end{tabular}
\caption{\label{tab:variE} Energy, degeneracy, and wave-vector
of the low-lying singlet states of the 12 and 16 site
symmetric clusters  
obtained for $J=1$ and $J'=0$ by exact diagonalization and 
by the variational approach with 0, 1 and 2 additional steps.
The second column is the energy of a single covering (no resonance between 
plaquettes). The number of plaquette coverings is equal to 
36 (84) for 12 (16) sites, while there are 462 (24024) singlets
in the full Hilbert space.
In the exact diagonalization spectrum, these states lie below the 
first multiplet ($E=-12.841$ and $-18.451$ for 12 and 16 sites respectively). 
}
\end{table}

To reach larger system sizes, we have decided to resort 
to variational calculations
in the spirit of the RVB approach to frustrated SU(2) magnets \cite{RVBsu2}.
There are significant differences though.
First, one needs at least four sites to make an SU(4) singlet. 
Now, among all possible
4-site clusters, the particular role of the
singlet plaquettes was pointed out by Li {\it et al.} \cite{lima}:
On a triangular lattice, each plaquette has a diagonal 
energy of $-(13/4)J$, which 
is much lower than the diagonal energy of the LRO state (which is 0). Thus the
singlet plaquettes are a good starting point for a variational calculation.
To implement the method,
we define a Hilbert space which is spanned by the different 
coverings $|\psi_j \rangle$ ($j=1,\dots,N_{\rm cov}$) of 
the finite size cluster with 
singlet plaquettes. 
The eigenenergies are then solution of the generalized eigenvalue
problem $\det(E O_{ij} - H_{ij})=0$,  where 
$O_{ij}=\langle \psi_i| \psi_j \rangle$ and 
$H_{ij}=\langle \psi_i| \mathcal{H} | \psi_j \rangle$.
A systematic way to improve the method is to double the Hilbert space by 
including the states $| \psi'_j \rangle= H |\psi_j \rangle$, which leads
to the
\begin{equation}
  \left| 
  \left(
     \begin{array}[c]{cc}
        H_{ij} & H^{(2)}_{ij} \\
        H^{(2)}_{ij} & H^{(3)}_{ij} 
     \end{array}
  \right)  
    -    
    E 
   \left( 
     \begin{array}[c]{cc}
        O_{ij} & H_{ij} \\
        H_{ij} & H^{(2)}_{ij} 
     \end{array}
  \right)  
\right| = 0
\end{equation} 
$2 N_{\rm cov} \times 2 N_{\rm cov}$ eigenvalue problem, where 
$H^{(n)}_{ij}=\langle \psi_i| \mathcal{H}^n | \psi_j \rangle$. This step 
can be repeated
to include matrix elements of further powers of $\mathcal{H}^n$.
The efficiency of the method is illustrated in Table.~\ref{tab:variE}.

The advantage of the variational approach is that we could perform
calculations for much larger clusters, up to 36 sites with 2 additional steps, 
48 sites with 1 additional step, and 64 sites with no additional step.
For clarity, we concentrate the discussion on two natural possibilities:
i) The singlet plaquettes form a pattern with long-range order,
the most natural candidate being then the static
covering (plaquette array) as in the square lattice \cite{RVBsu4} 
(See Fig.~\ref{fig:covering}(a)); ii) The plaquettes undergo
strong resonances between configurations that differ only locally,
the smallest pattern that allows such a resonance being the 
diamond depicted in Fig.~\ref{fig:covering}(b).

\begin{figure}
\includegraphics[width=5.0truecm]{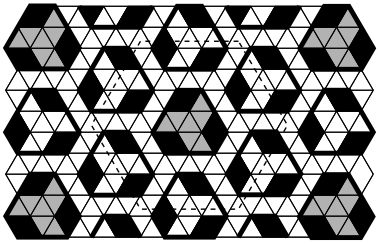}
\caption{\label{fig:gs} Dominant plaquette covering in the ground state
of the 48 site cluster (dashed line). The diamonds making up the 
new superlattice are shaded using different grays for the two 
patterns of Fig.~\ref{fig:covering}(b).}
\end{figure}

The first possibility is definitely not favoured by our variational
results: For both the 36-site and the 48-site clusters, the plaquette
covering which has the largest weight in the ground state wave function
is not the square covering, but a diamond covering (see Fig.~\ref{fig:gs}).
This fact is actually easy to understand: The matrix elements
corresponding to the resonance of Fig.~\ref{fig:covering}(b) dominate the
secular equation,
and they are completely suppressed in the square pattern of 
Fig.~\ref{fig:covering}(a).

The effect of these resonances on the ground state correlations is
a delicate issue. The ground states of the 36-site and 48-site 
clusters are consistent with a long-range pattern for the diamond
covering of the triangular lattice, but given the very small sizes 
from that point of view (the 36 and 48 site cluster can accommodate
3 and 4 diamonds respectively) this result is probably not significant. 
What is likely to be more relevant is the amplitude of the processes
that would enter a description in terms of an effective Hamiltonian. 
The two plaquette coverings inside a diamond can be described by
an Ising variable, and the resonance gives rise to a transverse field
of order $J/2$. The spectrum obtained for the 36-site and 48-site
clusters are then consistent with a very small potential energy term
with two and three body interactions. A detailed analysis of this
model lies beyond the scope of the present paper, but given the 
prominence of the kinetic energy term, it is very likely that the
system is in the equivalent of the disordered phase of the Ising 
model in a transverse field \cite{moessner}. 

In summary, this variational approach strongly suggests that the ground
state is a plaquette liquid with no four-site plaquette long-range order,
and with strong local resonances between the 
configurations of Fig.~\ref{fig:covering}(b). The presence of long-range
order associated to a specific pattern of diamond covering can neither
be ascertained nor excluded on the basis of the present results. We just
note that, if no longer range correlations are present, this plaquette
liquid is expected, as its SU(2) counterparts, to exhibit topological 
degeneracy \cite{topdeg}.

{\it Conclusion.}-- 
To conclude, we have shown that it is possible to order the SU(4) 
Heisenberg model for sufficiently large nearest neighbor repulsion,
without the need to introduce anisotropy, and we have identified 
a quantum phase transition around $J'/J=0.12$. For $J'=0$, which 
corresponds to a minimal model of LiNiO$_2$, we have shown that 
the ground state is a spin {\it and} orbital liquid, and we have 
shown that simple objects like dimers of plaquettes
do not develop long-range order. These results are consistent
with the absence of any kind of phase transition in LiNiO$_2$.
More generally, Mott insulators with orbital degeneracy
and the appropriate geometry emerge as potential candidates for 
completely order-free spin liquids with topological degeneracy.

\begin{acknowledgments}
We acknowledge the financial
support of the Hungarian OTKA T038162, D32689, Bolyai 118/99, and of the
Swiss National Fund.
\end{acknowledgments}

\end{document}